%% file: paper.tex
\documentclass[letterpaper,twocolumn,10pt]{article}
\usepackage{usenix-2020-09}

\usepackage[english]{babel}
\usepackage{blindtext}
\usepackage{epsfig}
\usepackage{tabularx}
\usepackage{graphicx} 
\usepackage{color}
\usepackage{xspace}
\usepackage{thumbpdf}
\usepackage{enumitem}
\usepackage{listings}
\usepackage{verbatim}
\usepackage{multirow}
\usepackage{booktabs}
\usepackage{colortbl}
\usepackage{subcaption}
\usepackage{amssymb,amsmath,amsfonts}
\usepackage{titlesec}
\usepackage{microtype}
\UseMicrotypeSet[protrusion]{basicmath} 
\usepackage{cleveref}

\usepackage{url}

\usepackage{breakurl}
\usepackage[breaklinks]{hyperref}

\crefformat{section}{\S#2#1#3} 
\crefformat{subsection}{\S#2#1#3}
\crefformat{subsubsection}{\S#2#1#3}

\usepackage[nomain,numberline,savewrites=true,acronym, debug]{glossaries}
\makeglossaries
\input{glossary}

\hypersetup{colorlinks=true, linkcolor=black, filecolor=blue, citecolor = black,
     urlcolor=blue}

\definecolor{coolblack}{rgb}{0.0, 0.18, 0.39}
\definecolor{darkmidnightblue}{rgb}{0.0, 0.2, 0.4}

\newcommand{\eg}{{\it e.g.}}
\newcommand{\ie}{{\it i.e.}}
\newcommand{\etal}{{\it et al.}}

\titlespacing\section{0pt}{12pt plus 4pt minus 2pt}{4pt plus 2pt minus 2pt}
\titlespacing\subsection{0pt}{8pt plus 4pt minus 2pt}{3pt plus 2pt minus 2pt}
\titlespacing\subsubsection{0pt}{4pt plus 4pt minus 2pt}{0pt plus 2pt minus 2pt}

\begin{document}
\sloppy

\title{Pretty Good Phone Privacy}

\author{
{\rm Paul Schmitt}\\
Princeton University\\
\and
{\rm Barath Raghavan}\\
University of Southern California  \\
} 

\maketitle 

\begin{abstract}
  \input{abstract}
\end{abstract}
\input{intro}

\input{case}

\input{measurement}

\input{questions}

\input{pgpp}

\input{analysis}
\input{discussion}\label{lastpage}

\bibliographystyle{plain}  
\interlinepenalty=10000
\bibliography{bib}

\section{Glossary}\label{sec:glossary}

\small \printglossary[type=acronym,style=altlist,title={},nogroupskip]
\label{endpage}

\end{document}

%% file: glossary.tex
\newacronym{aka}{AKA}{Authentication and Key Agreement. The process by which the UE and the AUSF exchange information by which they can each verify a secret key held by the other, and calculate keys to be used for ciphering and integrity protection of data transmitted between the UE and the network} 

\newacronym{eutran}{NG-RAN}{Next Generation Radio Access Network. Network that serves to connect UEs and gNodeBs}

\newacronym{sgw}{SMF}{Session Management Function. The session managment function supports session management and IP address allocation}

\newacronym{s1ap}{NGAP}{Next Generation Application Protocol. The signaling protocol used between the NG-RAN and the NGC}

\newacronym{pgw}{UPF}{User Plane Function. The gateway that provides global IP connectivity from the NGC.}

\newacronym{pgppgw}{PGPP-GW}{PGPP Gateway. A proposed gateway for PGPP that sits between the UPF and the global Internet. The PGPP-GW allows for billing without requiring the user's identity }

\newacronym{hss}{AUSF}{Authentication Server Function. The entity that holds subscription information to allow or deny access to the network}

\newacronym{mme}{AMF}{Access and Mobility Management Function. The control entity that manages signaling between the UE and the core network. AMF supports functions related to bearer and connection management and manages mobility between gNodeBs}

\newacronym{enb}{gNodeB}{Next Generation NodeB. The base station in 5G}

\newacronym{EPC}{NGC}{Next Generation Core. The core network in 5G. Main logical nodes of the NGC are the User Plane Function (UPF), Access and Mobility Management Function (AMF), the Session Management Function (SMF), and Authentication Server Function (AUSF)}

\newacronym{ue}{UE}{User Equipment. The mobile device which allows a user to access network services, connecting to the UTRAN or E-UTRAN via the radio interface. Commonly understood to be a mobile phone}

\newacronym{rnti}{RNTI}{Radio Network Temporary Identifier. A unique identifier for a UE in a given cell, used to connect over layer 2}

\newacronym{msisdn}{MSISDN}{Mobile Station International Subscriber Directory Number. The unique number that identifies a user of the global telecom system (e.g., a phone number) }

\newacronym{ims}{IMS}{IP Multimedia Subsystem. The entity that provides voice and messaging services for the network}

\newacronym{mno}{MNO}{Mobile Network Operator. A cellular service provider}

\newacronym{mvno}{MVNO}{Mobile Virtual Network Operator. A cellular operator that does not necessarily own its own spectrum or all of the network equipment it operates upon. MVNOs run on top of MNO networks}

\newacronym{pmvno}{PMVNO}{Private Mobile Virtual Network Operator. An MVNO that decouples user identity used for authentication and billing from network connectivity}

\newacronym{sim}{SIM}{Subscriber Identity Module. An entity that holds the IMSI, which uniquely identifies a subscriber. SIMs are used to authenticate a user to the network}

\newacronym{imsi}{IMSI}{International Mobile Subscriber Identity. A globally unique identifier associated with each mobile phone subscriber. It is stored in the SIM inside the phone and is sent by the phone to the network}

\newacronym{guti}{GUTI}{Globally Unique Temporary Identity. The GUTI is a temporary identifier that can be used in lieu of an IMSI to identify a subscriber to the core network}

\newacronym{imei}{IMEI}{International Mobile Equipment Identity. A globally unique, permanent device identifier which is allocated to each individual mobile device. It is set by the manufacturer}

\newacronym{eir}{EIR}{Equipment Identity Register. A database that stores IMEIs of devices in cellular systems. IMEIs can be white-listed, grey-listed or black-listed. The EIR allows a device's identity to be checked for blacklisting, (e.g., whether is has been reported stolen)}

\newacronym{ceir}{CEIR}{Centralized Equipment Identity Register. A globally centralized EIR database that can be used by multiple providers}

\newacronym{tau}{TAU}{Tracking Area Update. The procedure by which the UE updates the network as it moves to a new location (e.g., entered a new tracking area)}

\newacronym{tap}{TAP}{Transferred Account Procedure. A file detailing usage and wholesale charges due to roaming}

\newacronym{sqn}{SQN}{Sequence Number. A value stored at the AUSF and the SIM to maintain synchrony between the entities}

\newacronym{xres}{XRES}{Expected Response. A value generated by the core network used during the authentication procedure}

\newacronym{rand}{RAND}{A random nonce used during the authentication procedure}

\newacronym{ss7}{SS7}{Signaling System 7. The protocol standard used by entities on public switched telephone networks communicate with one another. SS7 is used to setup and tear down voice calls, deliver SMS, etc. SS7 has been largely replaced by Diameter in modern cellular standards}

\newacronym{diameter}{Diameter}{The authentication, authorization, and accounting protocol used by 4G/5G cellular networks. Diameter is used to enable roaming between modern cellular networks}

\newacronym{ecm}{ECM}{EPS Mobility Management. The set of routines used by the core network to manage mobility procedures such as attach and tracking area updates}

\newacronym{emm}{EMM}{EPS Connection Management. The set of routines used by the core network to manage connectivity between the UE and the NGC}

\newacronym{epsbearer}{EPS bearer}{EPS bearer. A virtual connection between a UE and the UPF}

\newacronym{tal}{TAL}{Tracking Area List. A list of tracking areas stored on the device that the device can enter without triggering a tracking area update}

\newacronym{ta}{TA}{Tracking Area. A tracking includes one or many gNodeBs. Typically, the UE can move freely within gNodeBs in a tracking area without notifying the AMF with a tracking area update}


%% file: abstract.tex
To receive service in today's cellular architecture, phones uniquely identify
themselves to towers and thus to operators.  This is now a cause of major
privacy violations, as operators sell and leak identity and location data
of hundreds of millions of mobile users.

In this paper, we take an end-to-end perspective on the cellular architecture
and find key points of decoupling that enable us to protect user identity and
location privacy with no changes to physical infrastructure, no added latency,
and no requirement of direct cooperation from existing operators.

We describe Pretty Good Phone Privacy (PGPP) and demonstrate how our modified
backend stack (\acrshort{EPC}) works with real phones to provide ordinary yet
privacy-preserving connectivity. We explore inherent privacy and efficiency
tradeoffs in a simulation of a large metropolitan region. We show how PGPP
maintains today's control overheads while significantly improving user identity
and location privacy.

%% file: intro.tex
\section{Introduction}
Cellular phone and data networks are an essential part of the global
communications infrastructure. In the United States, there are 124 cellular
subscriptions for every 100 people and the total number of cellular
subscriptions worldwide now stands at over 8.2 billion~\cite{worldbank}.
Unfortunately, today's cellular architecture embeds privacy assumptions of a
bygone era. In decades past, providers were highly regulated and centralized,
few users had mobile devices, and data broker ecosystems were undeveloped. As a
result, except for law enforcement access to phone records, user privacy was
generally preserved. Protocols that underpin cellular communication embed an
assumption of trusted hardware and infrastructure~\cite{3gpp.23.401}, and
specifications for cellular backend infrastructure contain few formal
prescriptions for preserving user data privacy. The result is that the locations
of all users are constantly tracked as they simply carry a phone in their
pocket, \textit{without even using it}.

Much has been made of privacy enhancements in recent cellular standards (\eg,
5G), but such changes do nothing to prevent cellular carriers from tracking user
locations. Worse still, the 5G push toward small cells results in much
finer-grained location information, and thus tracking, than previous
generations.

\paragraph{Privacy violations by carriers.} In recent years it has been
extensively reported that mobile carriers have been routinely selling and
leaking mobile location data and call metadata of hundreds of millions of
users~\cite{zdnet,nytimes-cell,adage,vice,vice2}. Unfortunately for users, this
behavior by the operators appears to have been legal, and has left mobile users
without a means of recourse due to the confluence of a deregulated industry,
high mobile use, and the proliferation of data brokers in the landscape. As a
result, in many countries every mobile user can be physically located by anyone
with a few dollars to spend. This privacy loss is ongoing and is independent of
leakage by apps that users choose to install on their phones (which is a related
but orthogonal issue).

While this major privacy issue has long been present in the architecture, the
practical reality of the problem and lack of technical countermeasures against
bulk surveillance is beyond what was known before. However there is a
fundamental technical challenge at the root of this problem: even if steps were
taken to limit the sale or disclosure of user data, such as by passing
legislation, the cellular architecture generally and operators specifically
would still seemingly need to know where users are located in order to provide
connectivity. Thus, as things stand, users must trust that cellular network
operators will do the right thing with respect to privacy despite not having
done so to date.

\paragraph{Architectural, deployable solution.} We identify points of
decoupling in the cellular architecture to protect user privacy in a way that
is immediately deployable. In this, we are aided by the industry-wide shift
toward software-based cellular cores. Whereas prior generations of cellular
networks ran on highly-specific hardware, many modern cellular core functions
are run in software, making it more amenable to key changes. 

In our approach, users are protected against location tracking, even by their
own carrier. We decouple network connectivity from authentication and billing,
which allows the carrier to run Next Generation Core (\acrshort{EPC}) services
that are unaware of the identity or location of their users but while still
authenticating them for network use. Our architectural change allows us to
nullify the value of the user's \acrshort{imsi}, an often targeted identifier in
the cellular ecosystem, as a unique identifier. We shift authentication and
billing functionality to outside of the cellular core and separate traditional
cellular credentials from credentials used to gain global
connectivity.

Since it will take time for infrastructure and legislation to change, our work
is explicitly \emph{not} clean slate. We anticipate that our solution is most
likely to be deployed by Mobile Virtual Network Operators (\acrshort{mvno}s),
where the \acrshort{mvno} operates the core (\acrshort{EPC}) while the base
stations (\acrshort{enb}s) are operated by a Mobile Network Operator
(\acrshort{mno}). This presents us with architectural independence as the
\acrshort{mvno} can alter its core functionality, so long as the \acrshort{EPC}
conforms to LTE / 5G standards. While it is not strictly necessary for PGPP to be
adopted by an \acrshort{mvno}, we assume that existing industry players (\eg,
\acrshort{mno}s) are unlikely to adopt new technologies or have an interest in
preserving user privacy unless legal remedies are instituted. As a result, we
consider how privacy can be added on top of today's mobile infrastructure by new
industry entrants. 



%

\paragraph{Contributions.} We describe our prototype implementation,
Pretty Good Phone Privacy (PGPP). In doing so, we examine several key challenges
in achieving privacy in today's cell architecture. In particular, we consider:
1) which personal identifiers are stored and transmitted within the cellular
infrastructure; 2) which core network entities have visibility into them (and
how this can be mitigated); 3) which entities have the ability to provide
privacy and with what guarantees; and 4) how we can provide privacy while
maintaining compatibility with today's infrastructure and without requiring the
cooperation of established providers. 


We show PGPP's impact on control traffic and on user anonymity. We show that by
altering the network coverage map we are able to gain control traffic headroom
compared with today's networks; we then consume that headroom in exchange for
improved anonymity. We analyze the privacy improvements against a variety of
common cellular attacks, including those based on bulk surveillance as well as
targeted attacks. We find that PGPP significantly increases anonymity where there is none today. We
find that an example PGPP network is able to increase the geographic area that
an attacker could believe a victim to be within by \textasciitilde 1,200\% with
little change in control load.\\[1ex]
\noindent Our contributions are as follows:
\begin{itemize}
    \item We design a new architecture that decouples connectivity
    from authentication and billing functionality, allowing us to alter the
    identifiers used to gain connectivity (\cref{sec:imsis}) and enable
    PGPP-based operators to continue to authenticate and bill users
    (\cref{sec:auth}) without identifying them.
    \item We adapt existing mechanisms to grow control traffic broadcast
    domains, thus enhancing user location privacy while maintaining backwards
    compatibility (\cref{sec:locationprivacy}).
    \item We quantify the impacts of PGPP on both user privacy and network
    control traffic through simulation (\cref{sec:simulation}) and demonstrate
    PGPP's feasibility in a lab testbed.
\end{itemize}

%% file: case.tex
\section{Background}

Here we provide a brief overview of the cellular architecture and describe the
inherent privacy challenges. For simplicity we focus on 5G, though the
fundamental challenges also exist in legacy standards.

\subsection{Cellular architecture overview}\label{sec:background} 
\begin{figure}[t]
\centering
\includegraphics[width=0.95\columnwidth]{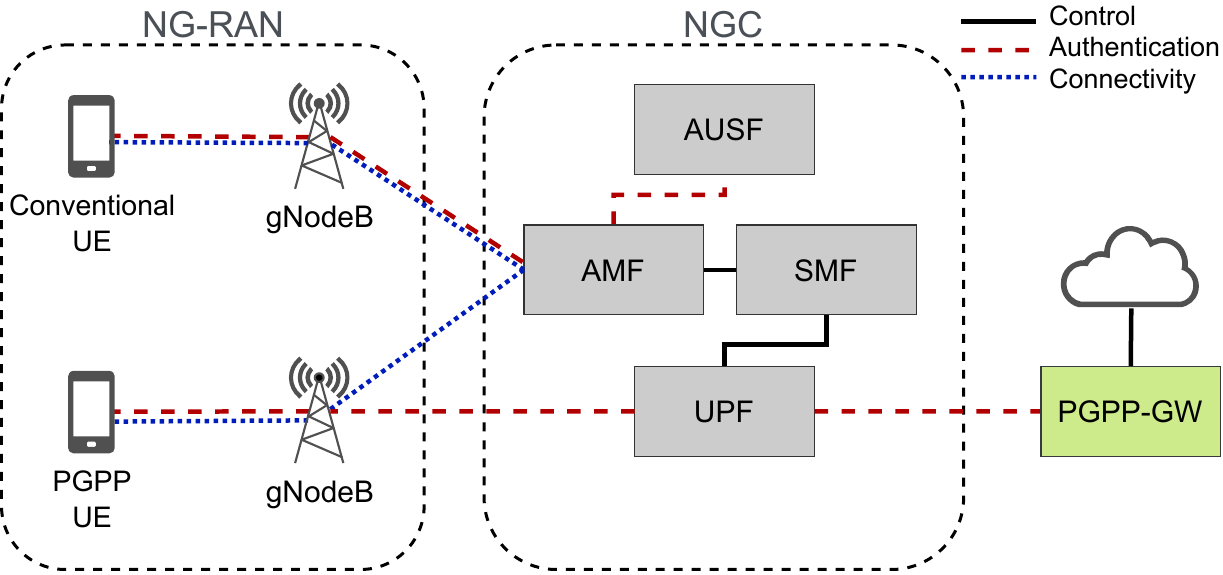}
\caption{Simplified 5G architecture with and without PGPP. PGPP decouples authentication and connectivity credentials and shifts authentication to a new, external entity, the PGPP-GW. Details of the PGPP-GW are found in~\cref{sec:auth}.}
\label{fig:lte_arch}
\vspace{-3mm}
\end{figure}

The 5G
architecture can be divided into two areas: the Next Generation Radio Access
Network (\acrshort{eutran}), which is responsible for radio access; and the Next
Generation Core (\acrshort{EPC}), which includes the entities responsible for
authentication and connectivity to the network core. Figure~\ref{fig:lte_arch}
shows a simplified architecture for both conventional cellular as well as with
PGPP. PGPP moves authentication and billing to a new entity, the PGPP-GW, that
is external to the \acrshort{EPC}. We detail PGPP's specific changes
in~\cref{sec:pgpp}. We include a glossary of cellular terms in
Appendix~\ref{sec:glossary}.

\textit{\acrshort{eutran}.} The \acrshort{eutran} is the network that
facilitates connectivity between user devices (\acrshort{ue}s)---commonly a cell
phone with a \acrshort{sim} card installed---and the serving base station
(\acrshort{enb}). The \acrshort{eutran} is responsible for providing
\acrshort{ue}s a means of connecting to the \acrshort{EPC} via \acrshort{enb}s.

\textit{\acrshort{EPC}.} The \acrshort{EPC} is the core of the 5G cellular
network and includes entities that provide authentication, billing, voice, SMS,
and data connectivity. The \acrshort{EPC} entities relevant to our discussion
are the Access and Mobility Management Function (\acrshort{mme}), the
Authentication Server Function (\acrshort{hss}), the Session Management Function
(\acrshort{sgw}), and the User Plane Function (\acrshort{pgw}). The
\acrshort{mme} is the main point of contact for a \acrshort{ue} and is
responsible for orchestrating mobility and connectivity. \acrshort{ue}s
authenticate to the network by sending an identifier that is stored in the
\acrshort{sim} to the \acrshort{mme}. The \acrshort{hss} is then queried to
verify that the \acrshort{ue} is a valid subscriber. Once the \acrshort{ue} is
authenticated, the \acrshort{mme} assigns the \acrshort{ue} to an \acrshort{sgw}
and \acrshort{pgw}, which offer an IP address and connectivity to the Internet.
Note that 5G networks can include many copies of these entities and contain many
more entities; however, for the purposes of our discussion this simplified model
suffices.

\textit{\acrshort{mvno}s.} We design our solution to be implemented by a Mobile
Virtual Network Operator (\acrshort{mvno}). \acrshort{mvno}s are virtual in that
they offer cellular service without owning the infrastructure itself. Rather,
\acrshort{mvno}s pay to share capacity on the infrastructure that an underlying
carrier operates. \acrshort{mvno}s can choose whether they wish to operate their
own core entities such as the \acrshort{mme}, \acrshort{hss}, and
\acrshort{pgw}, which is the type of operation we propose. \acrshort{mvno}s that
run their own core network are often called ``full'' \acrshort{mvno}s.
Critically, our architecture is now feasible as the industry moves toward
``whitebox'' \acrshort{enb}s that connect to a central office that is a
datacenter with virtualized \acrshort{EPC} services, as in the Open Networking
Foundation's M-CORD project~\cite{m-cord}. Recent work has shown that dramatic
performance gains are possible using such newer architectures~\cite{PEPC,klein}.

\subsection{Privacy in the cellular architecture}\label{sec:privbackground}
Maintaining user privacy is challenging in cellular networks, both past and
present as it is not a primary goal of the architecture. In order to
authenticate users for access and billing purposes, networks use globally unique
client identifiers. Likewise, the cellular infrastructure itself must always
``know'' the location of a user in order to minimize latency when providing
connectivity. We briefly discuss cellular identifiers as well as location
information available from the perspective of the cell network in this section.
We use acronyms from the 5G architecture as it is the newest standard; however,
similar entities exist in all generations (2G, 3G, 4G LTE).

\paragraph{User and device identifiers.}\label{sec:ids} There are multiple
identifiers that can be used to associate network usage with a given subscriber.
Identifiers can be assigned by various actors in the ecosystem, they can vary in
degree of permanence, and they can be globally unique across all cellular
operators or they can be locally unique within a given network.
Table~\ref{t:identifiers} shows these identifiers, their allocators, and their
permanence.
\begin{table}[t]
\centering
\scalebox{1.0}{
\begin{tabular}{|l|l|l|}
\hline
\textbf{Identifier}  & \textbf{Allocator} & \textbf{Duration} \\ \hline\hline
\acrshort{imsi}                 & Operator           & Permanent         \\ \hline
\acrshort{guti}                 & \acrshort{mme}                & Temporary         \\ \hline
IP Address (static)  & Operator           & Permanent         \\ \hline
IP Address (dynamic) & \acrshort{pgw}               & Temporary         \\ \hline
\acrshort{rnti}                 & \acrshort{enb}                & Temporary         \\ \hline
\end{tabular}}
\caption{User identifiers in LTE.}
\vskip -1em
\label{t:identifiers}
\end{table}

The International Mobile Subscriber Identity (\acrshort{imsi}) is the identifier
used to gain access to the network when a phone (\acrshort{ue}) performs initial
attachment. The \acrshort{imsi} is globally unique, permanent, and is stored on
the \acrshort{sim} card. Carriers maintain a \acrshort{hss} database containing
the list of \acrshort{imsi}s that are provisioned for use on the network and
subscription details for each. Because the \acrshort{imsi} is globally unique
and permanent, it is seen as a high-value target for those who wish to surveil
cellular users. For example, in recent years there has been a rise of cell-site
simulators, also known as \acrshort{imsi} catchers. These devices offer what
appears to be a legitimate base station (\acrshort{enb}) signal. Since
\acrshort{ue} baseband radios are na\"ive and automatically connect to the
strongest signal, they will attempt to attach to the \acrshort{imsi} catcher and
offer their \acrshort{imsi}. \acrshort{imsi} catchers have been used extensively
by law enforcement as well as nation-state adversaries to identify and eavesdrop
on cellular users~\cite{paget2010practical}. 

Given the \acrshort{imsi}'s importance and sensitivity, temporary identifiers
are often used instead. The Globally Unique Temporary Identifier
(\acrshort{guti}) can be thought of as a temporary replacement for an
\acrshort{imsi}. Once a phone attaches to the network, the Access and Mobility
Management Function (\acrshort{mme}) generates a \acrshort{guti} value that is
sent to the \acrshort{ue}, which stores the value. The \acrshort{ue} uses the
\acrshort{guti} rather than the \acrshort{imsi} when it attaches to the network
in the future. The \acrshort{guti} can be changed by the \acrshort{mme}
periodically. Prior work recently found that \acrshort{guti}s are often
predictable with consistent patterns, thus offering little
privacy~\cite{hong2018guti}, but this can be remedied with a lightweight fix
that we expect will be used going forward.


The 5G network is IP-based, meaning \acrshort{ue}s must be given IP addresses in
order to connect. IPs can be either statically or dynamically assigned to
\acrshort{ue}s. Statically assigned IPs are stored in a backend core database.
During the attach procedure, the \acrshort{mme} retrieves the static IP address
assigned to the \acrshort{ue} from the backend. Conversely, dynamic addresses
are assigned by the \acrshort{sgw} when the \acrshort{ue} attaches. Providers
can associate a user with an IP address in the network by monitoring traffic at
the \acrshort{pgw}, which offers a convenient location to place a network tap.

\begin{figure*}[t]
	\begin{subfigure}[t]{0.31\textwidth}
		\centering
		\includegraphics[width=\columnwidth]{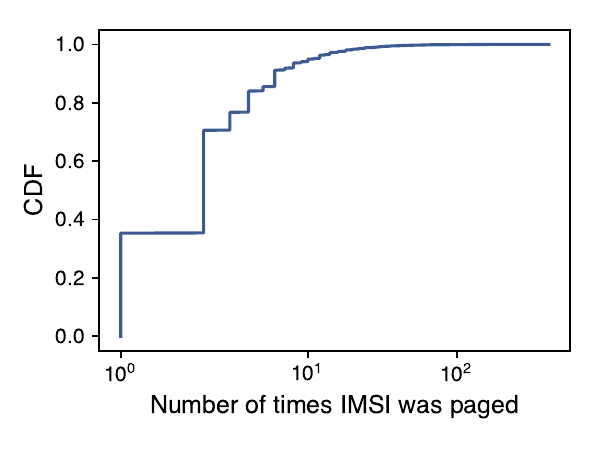}
		\caption{IMSI page counts.}
		\label{fig:zaataripages}
    \end{subfigure}
    \hfill
    \begin{subfigure}[t]{0.31\textwidth}
		\centering
		\includegraphics[width=\columnwidth]{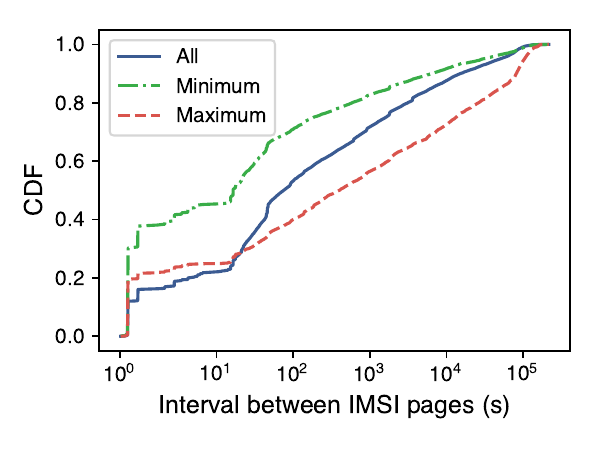}
		\caption{Intervals between pages.}
		\label{fig:zaatariintervals}
    \end{subfigure}
    \hfill
    \begin{subfigure}[t]{0.30\textwidth}
		\centering
		\includegraphics[width=\columnwidth]{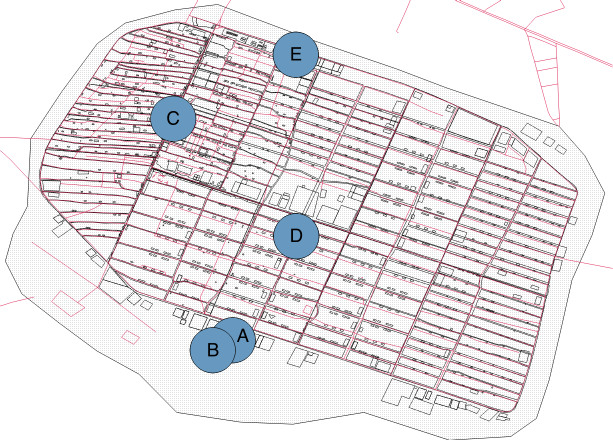}
		\caption{User locations over time.}
		\label{fig:zaatarimap}
    \end{subfigure}
    \caption{Analysis of IMSI broadcasts based on cellular traces captured in measurement study.}
\end{figure*}

In order to connect with the \acrshort{enb} over the \acrshort{eutran},
\acrshort{ue}'s must be assigned radio resources at layer 2, including a
temporary unique identifier, the \acrshort{rnti}. Prior work has shown that
layer 2 information used on the \acrshort{eutran} can be used to link
\acrshort{rnti}s with temporary identifiers at higher layers (\eg,
\acrshort{guti}s) provided the attacker knows the \acrshort{guti}
beforehand~\cite{8835335}. This attack is specific to the coverage area of a
single cell, and can be mitigated by changing the \acrshort{guti} frequently,
as discussed in~\cite{hong2018guti}.

\paragraph{User location information.}\label{sec:location} Cellular networks
maintain knowledge of the physical location of each \acrshort{ue}. Location
information is necessary to support mobility and to quickly find the
\acrshort{ue} when there is an incoming call, SMS, or data for a user. The
mechanism used to locate a \acrshort{ue} is known as ``paging'' and it relies on
logical groupings of similarly located \acrshort{enb}'s known as ``tracking
areas'' (\acrshort{ta}s). Each \acrshort{enb} is assigned to a single
\acrshort{ta}. \acrshort{ta}s can be thought of as broadcast domains for paging
traffic. If there is incoming data for an idle \acrshort{ue}, the paging
procedure is used, where the network sends a paging message to all
\acrshort{enb}s in the user's last-known \acrshort{ta}. Prior work has shown
that the paging mechanism can be leveraged by attackers that know an
identifier of the victim (\eg, phone number, WhatsApp ID) to generate paging
messages intended for the victim, which enables an unprivileged attacker to
identify a specific user's location~\cite{kune2012location}. From an external
perspective, the vantage point of remote servers on the web can also be
leveraged to localize mobile users given timing information from applications on
their devices~\cite{8413110}.

Cellular operators often store location metadata for subscriber, giving them the
ability to trace user movement and location history. This bulk
surveillance mechanism has been used to establish a user's past location by law
enforcement~\cite{carpenter_v_united_states_2018}.

%% file: measurement.tex
\section{The need for privacy enhancements}\label{sec:zaatarimeasurement} 

In this section we demonstrate the privacy leakage that exists in today's
cellular architecture by conducting a measurement study while acting as a
relatively weak attacker in a real-world environment. Recall
from~\cref{sec:privbackground} that the \acrshort{imsi} is a globally unique,
permanent identifier. Unfortunately for user privacy, the traditional cellular
architecture uses \acrshort{imsi}s for authentication and billing, as well as
providing connectivity, causing the \acrshort{imsi} to be transmitted for
multiple reasons.

Because of its importance and permanence, the \acrshort{imsi} is seen as a
high-value target for those who wish to surveil cellular users. For example, in
recent years there has been a proliferation of cell-site simulators, also known
as \acrshort{imsi} catchers. These devices offer what appears to be a legitimate
base station (\acrshort{enb}) signal. Since \acrshort{ue} baseband radios are
na\"ive and automatically connect to the strongest signal, they attempt to
attach to the \acrshort{imsi} catcher and offer their \acrshort{imsi}.
\acrshort{imsi} catchers have been used extensively by law enforcement and
state-level surveillance agencies, with and without warrants, to identify,
track, and eavesdrop on cellular users~\cite{paget2010practical}.

\paragraph{Dataset.}\label{sec:zaatari} We analyze a dataset of cellular
broadcast traces that our team gathered in a small, densely populated area with
roughly 80,000 residents over the course of several days in 2015. The traces
include messages that were sent on broadcast channels in plaintext for three
cellular providers that offer service in the area. Traces were captured using
software defined radios and mobile phones. The trace dataset provides a vantage
point that is akin to an \acrshort{imsi} catcher.\footnote{Trace collection
methodology and analysis received IRB approval; extraneous details omitted for
blind review.}

\paragraph{\acrshort{imsi}s are often broadcast in-the-clear.} We
discover that, while the architecture is designed to largely use temporary
\acrshort{guti}s once \acrshort{ue}s are connected, \acrshort{imsi}s are often
present in paging messages. Overall we see 588,921 total paging messages, with
38,917 containing \acrshort{imsi}s (6.6\% of all pages). Of those messages we
see 11,873 unique \acrshort{imsi}s. We track the number of times each individual
\acrshort{imsi} was paged and plot a CDF in Figure~\ref{fig:zaataripages}. As
shown, more than 60\% of \acrshort{imsi}s were paged more than once in the
traces. Note that we count multiple pages seen within one second as a single
page. Given this network behavior, even a passive eavesdropper could learn the
permanent identifiers of nearby users.

\paragraph{\acrshort{imsi}s can be tracked over time.} Given that
\acrshort{imsi}s are regularly broadcast, an eavesdropper can track the presence
or absence of users over time. We investigate the intervals between pages
containing individual \acrshort{imsi}s. In Figure~\ref{fig:zaatariintervals} we
plot a CDF of intervals (greater than one second) between subsequent pages of
individual \acrshort{imsi}s. Overall, we see that \acrshort{imsi}s are
repeatedly broadcast over time, even though the design of the architecture
should dictate that \acrshort{imsi}s should be used sparingly in favor of
temporary \acrshort{guti}s.
%

\paragraph{Individuals can be tracked over time.} Given that we can track
\acrshort{imsi}s over time, a passive attacker can track individuals' movements.
Figure~\ref{fig:zaatarimap} shows locations of base stations that broadcast the
\acrshort{imsi} for a single user in the traces. As shown, we saw the user in
multiple locations over the course of two days. Location A was recorded at 10am
on a Monday; location B was thirty minutes later. The user connected to a base
station at location C at noon that same day. Locations D and E were recorded the
following day at noon and 1:30pm, respectively. From this we see that a passive
observer unaffiliated with a cellular carrier can, over time, record the
presence and location of nearby users. This attacker is weak, with a relatively
small vantage point. In reality, carriers \textit{can and do} maintain this
information for \textit{all} of their users.

%% file: questions.tex
\section{Scope}
We believe that many designs are possible to increase privacy in
mobile networks, and no architecture, today or in the future, is likely to
provide perfect privacy.  Nevertheless, below we discuss various properties
that PGPP strives to achieve.

Prior work examined the security vulnerabilities in modern cell
networks~\cite{lteinspector,practicalattackslte,kune2012location} and revealed a number of flaws in the architecture itself. In addition, data brokers
and major operators alike have taken advantage of the cellular architecture's
vulnerabilities to profit off of revealing sensitive user data. We believe
mobile networks should aim to, at a minimum, provide one or both of the
following privacy properties:

\begin{itemize}[noitemsep]
\item \textit{Identity privacy.}  A network can aim to protect users' identity. Networks---as well as third party attackers---identify users through
\acrshort{imsi}s, which are intended to be uniquely identifying.
\item \textit{Location privacy.}  A network can aim to protect
information about the whereabouts of a phone.
\end{itemize}

Naturally, these privacy properties do not exist in isolation; they intersect in
critical ways. For example, attackers often aim to learn not only who a user is
but where a specific user is currently located, or where a user was when a
specific call was made.  Also, the definition of an attacker or adversary is a
complex one, and depending on context may include individuals aiming to steal
user data, mobile carriers and data brokers looking to profit off of user data,
governments seeking to perform bulk surveillance, law enforcement seeking to
monitor a user with or without due process, and many others.  Due to context
dependence, we do not expect all privacy-focused mobile networks to make the
same choice of tradeoffs.

\subsection{Cellular privacy threat model}\label{sec:attacks} Given the above
discussion, we distinguish between bulk and targeted data collection. We define
bulk collection to be the collection of information from existing cellular
architecture traffic without the introduction of attack traffic; thus, bulk
collection is passive. Bulk attacks commonly target user identities (\eg,
\acrshort{imsi}s). PGPP's core aim is to protect against bulk attacks. Targeted attacks are active and require injection of traffic
to attack specific targets. Targeted attacks are often aimed at discovering a
victim's location. We also delineate attacks by the adversary's capabilities, as
they may have visibility into an entire network (global) versus, for an
unprivileged attacker, some smaller subset of a network's infrastructure
(local). Table~\ref{tab:attacks} gives the taxonomy of attacks.

Carriers and governments are the most common \textbf{global-bulk} attackers.
Such bulk surveillance is commonplace in cellular networks, and has been at the
center of recent lawsuits and privacy concerns. 
Attacks that employ \acrshort{imsi} catchers or passively listen to broadcasts using
software-defined radios are considered \textbf{local-bulk}. Here, an \acrshort{imsi}
catcher is only able to monitor phones that connect directly to it, so its
visibility is limited to its radio range. Similarly, SDR-based passive snooping
(as in the example in~\cref{sec:zaatarimeasurement}) is only able to monitor
nearby base stations and will miss portions of the network. We design PGPP with
a primary focus on thwarting bulk attacks by nullifying the value of
\acrshort{imsi}s (\cref{sec:imsis}).

\begin{table}[t]
    \centering
    \small
    \resizebox{\columnwidth}{!}{
    \begin{tabular}{ll|l|l|}
    \cline{3-4}
                                                               &                 &
    \multicolumn{2}{c|}{\textbf{Attack type}}
    \\ \cline{3-4} \multicolumn{1}{l}{}                                      &
    & \textbf{Bulk}
    & \textbf{Targeted} \\ \hline
    \multicolumn{1}{|l|}{\multirow{2}{*}{\rotatebox[origin=c]{90}{\textbf{Visibility\hspace{-0.3em}}}}} & \textbf{Global} & \begin{tabular}[c]{@{}l@{}}Carrier logs~\cite{zdnet,adage,vice,vice2} / \\ Government Surveillance~\cite{carpenter_v_united_states_2018}\end{tabular} &  Carrier Paging \\ \cline{2-4} 
    \multicolumn{1}{|l|}{} & \textbf{Local}  & \begin{tabular}[c]{@{}l@{}}SDR~\cite{7146071, van2016effectiveness, mjolsnes2017easy} / \\ IMSI Catcher~\cite{paget2010practical, joachim2003method}\end{tabular} & Paging attack~\cite{hussain2019privacy, kune2012location}\\ \hline
    \end{tabular}}
    \caption{Common cellular attacks.}
    \label{tab:attacks}
    \vspace{-3mm}
    \end{table}

\textbf{Local-targeted} attacks can be carried out by ordinary users by
generating traffic that causes a network to page a victim (\eg, phone call to the victim). As local-targeted attackers do not have
visibility into the entire network, they must rely upon knowledge of the
geographic area that is encompassed by a tracking area. Due to the prevalence of
such attacks, as an enhancement, an operator can provide functionality, in
cooperation with the user, that reduces the efficacy of local-targeted attacks
through the use of \acrshort{tal}s~(\cref{sec:locationprivacy}).

\textbf{Global-targeted} attacks represent a very powerful attacker who can
actively probe a victim while having global visibility of the network. We envision
defenses against such attacks would require fundamental changes to to communication models. 
PGPP does not
mitigate global-targeted attacks as we focus on immediately deployable
solutions; we leave this to future work.

\subsection{Aims}
Next we discuss the aims of PGPP by considering several common questions that
arise. 

\textit{\textbf{What sort of privacy does PGPP provide?}} As its name suggests,
PGPP aims to provide ``pretty good'' privacy; we don't believe there is a
solution that provides perfect privacy, causes no service changes (\ie, does not
increase latency), and is incrementally deployable on today's cellular networks.
The main focus is to offer privacy against global-bulk surveillance of mobility
and location, a practice by carriers that is widespread and pernicious. We
thwart this via eliminating the \acrshort{imsi} as an individual identifier and
decoupling the authentication and connectivity mechanisms in the cellular
architecture. 


\textit{\textbf{Isn't 5G more secure than legacy generations?}} \label{sec:5g}
The 5G standard includes enhancements focused on user privacy and system
performance over legacy cellular generations. However, the enhancements do not
offer location privacy benefits from the carriers.


\textit{Encrypted \acrshort{imsi}s.} 5G includes the addition of encrypted
\acrshort{imsi}s, where public key cryptography, along with ephemeral keys
generated on the \acrshort{sim}, is used to encrypt the \acrshort{imsi} when
sending it to the network. This protects user \acrshort{imsi}s from
eavesdroppers. However, encrypted \acrshort{imsi}s do not prevent the cellular
provider \textit{itself} from knowing the user's identity. An analogy for
encrypted \acrshort{imsi}s can be found in DNS over HTTPS (DoH): eavesdroppers
cannot see unencrypted traffic, yet the endpoints (the DNS resolver for DoH, the
cellular core in 5G) still can. The goal of this work is to not only thwart
local-bulk attacks, but also protect user privacy from mobile operators that
would otherwise violate it (\ie, global-bulk attacks).

\textit{Small cell location privacy.} The 5G standard strives for reduced
latencies as well as much higher data throughputs. This necessitates the use of
cells that cover smaller areas in higher frequency spectrum in order to overcome
interference compared with previous cellular generations that used macrocells to
provide coverage to large areas. A (likely unintended) byproduct of 5G's use of
smaller cells is a dramatic {\em reduction} in location privacy for users. As
the 5G network provider maintains state pertaining to the location in the
network for a given user for the purposes of paging, smaller cells result in the
operator, or attacker, knowing user locations at a much higher precision
compared with previous generations.

\textit{\textbf{What about active | traffic analysis | signaling attacks?}}
While active, targeted attacks aren’t our main focus, we improve privacy in the
face of them by leveraging \acrshort{tal}s to increase and randomize the
broadcast domain for paging traffic, making it more difficult for attackers to
know where a victim is located (analyzed in~\cref{sec:localtargeted}).
Further, the goal of many active attacks is to learn users'
\acrshort{imsi}s, and our nullification of \acrshort{imsi}s renders such attacks
meaningless.

An attacker with a tap at the network edge could use traffic analysis attacks to
reduce user privacy. We largely view this as out of scope as users can tunnel
traffic and use other means to hide their data usage patterns.

Cellular networks rely on signaling protocols such as Signaling System 7
(\acrshort{ss7}) and \acrshort{diameter} when managing mobility as well as voice
and SMS setup and teardown. These protocols enable interoperability between carriers needed for roaming and connectivity
across carriers. Unfortunately, these protocols were designed with inherent
trust in the network players, and have thus been used to reduce user privacy and
disrupt
connectivity~\cite{lorenz2001securing,sengar2006ss7,engel2008locating,holtmans2016detach,sonar}.
We design PGPP for 4G/5G data only, which renders legacy \acrshort{ss7}
compatibility moot. Our PGPP design expects users to use outside messaging
services rather than an in-NGC \acrshort{ims} system. 

\textit{\textbf{Can PGPP support roaming?}} Yes. While we envision that many
PGPP users would explicitly not wish to roam, as roaming partners may not
provide privacy guarantees, roaming is possible using a \acrshort{diameter} edge
agent that only allows for home routed roaming, forcing traffic to route from
the visited network's \acrshort{sgw} back to the PGPP operator's \acrshort{pgw},
rather than local breakout due to our authentication mechanism
(\cref{sec:auth}). Roaming, and international roaming in particular, adds
billing complexities for the PGPP operator. Typically, the visited network
collects call data records for each roaming user on its network and calculates
the wholesale charges payable by the home network. The visited network then
sends a Transferred Account Procedure (\acrshort{tap}) file to the home network via a data
clearing house. The home network then pays the visited network. In PGPP, the
individual identity of the user that roamed is not known, yet the PGPP operator
remains able to pay the appropriate fees to visited networks.

\textit{\textbf{How does PGPP protect user privacy for voice or text service?}}
Out of the box, PGPP doesn't provide protection for such service. Instead, PGPP
aims provide privacy from the cellular architecture itself, and in doing so
users are free to use a third party VoIP provider (in which case the phone will
operate identically to a normal phone for telephony service from a user's
perspective) or use recent systems by Lazar et al.~\cite{yodel, karaoke} that
provide strong metadata privacy guarantees for communications, or similar
systems such as~\cite{199303,
vandenHooff:2015:VSP:2815400.2815417,corrigan2010dissent,Corrigan-Gibbs:2015:RAM:2867539.2867658}.
We view PGPP as complementary to such systems.

\textit{\textbf{How does PGPP protect users against leaky apps?}} PGPP doesn't,
as it is about providing protection in the cellular infrastructure. Even without
leaky apps, users can always intentionally or inadvertently reveal their
identity and location. Leaky apps make this worse as they collect and,
sometimes, divulge sensitive user information. We see PGPP as
complementary to work that has targeted privacy in mobile app ecosystems.
Further, apps are not as fundamental as connectivity---users can choose whether
to install and run a leaky app, and can constrain app permissions. However,
phones are, by their nature, always connected to carrier networks, and those
very networks have been selling user data to third parties.

\textit{\textbf{If users can't be identified by carriers, how can carriers still
make money?}} We introduce PGPP tokens in~\cref{sec:auth} as a mechanism for a
PGPP operator to charge customers while protecting user anonymity.

\textit{\textbf{Can't phone hardware be tracked as well?}} Phones have an
International Mobile Equipment Identity (\acrshort{imei}). The \acrshort{imei}
is assigned to the hardware by the manufacturer and identifies the manufacturer,
model, and serial number of a given device. Some operators keep an
\acrshort{imei} database to check whether a device has been reported as stolen,
known as an equipment identity register (\acrshort{eir}); \acrshort{imei}s in
the database are blacklisted. 


For many devices, the \acrshort{imei} can be changed through software, often
without root access. We envision a PGPP \acrshort{mvno} would allow for
subscribers to present their unchanged device \acrshort{imei}, giving the PGPP
operator the opportunity to check against a \acrshort{eir} to verify the phone
has not been reported as stolen. At that point, the \acrshort{imei} could be
reprogrammed to a single value, similar to our changes to the \acrshort{imsi}.
Note that different jurisdictions have different rules about whether, how, and
by whom an \acrshort{imei} can be changed, so only in some cases \acrshort{imei}
changes require cooperation with the \acrshort{mvno}.

\textit{\textbf{Is PGPP legal?}} Legality varies by jurisdiction. For example,
U.S. law (CALEA~\cite{calea}), requires providers to offer lawful interception
of voice and SMS traffic. A PGPP-based carrier is data-only, with voice and
messaging provided by third parties. CALEA requires the provider to offer
content of communication data at the \acrshort{pgw}, \eg, raw
(likely-encrypted) network traffic. This is supported by PGPP.

%% file: pgpp.tex
\section{Design}\label{sec:pgpp} 

In this section we describe the mechanisms PGPP employs to increase user
identity and location privacy. Ultimately, PGPP's design choices appear obvious
in retrospect. We believe its simplicity is an asset, as PGPP is compatible with
existing networks and immediately deployable. 

In order to provide identity privacy against bulk attacks, we nullify the value
of the \acrshort{imsi}, as it is the most common target
identifier for attackers. In our design, we choose to set all PGPP user
\acrshort{imsi}s to an identical value to break the link between \acrshort{imsi}
and individual users. This change requires a fundamental shift in the
architecture, as \acrshort{imsi}s are currently used for connectivity as well as
authentication, billing, and voice/SMS routing. We design a new cellular entity
for billing and authentication that preserves identity privacy. Fortunately, the
industry push for software-based \acrshort{EPC}s makes our architecture
feasible. We describe the architecture in~\cref{sec:imsis}.

To provide location privacy from targeted attacks, PGPP leverages an existing
mechanism (\acrshort{tal}s) in the cellular specification in order to grow the
broadcast domain for control traffic (\cref{sec:locationprivacy}). By changing
the broadcast domain for every user, the potential location of a victim is
broadened from the attacker's vantage point.

\subsection{User identity privacy}\label{sec:imsis} As discussed
in~\cref{sec:ids}, \acrshort{imsi}s are globally unique, permanent identifiers.
As such, they are routinely targeted by attackers, both legal and illegal. In
this section we re-architect the network in order to thwart {\em bulk} attacks
introduced in~\cref{sec:attacks} that are based on identifying individuals via
\acrshort{imsi}.


We decouple back-end connectivity from the authentication procedure that
normally occurs at the \acrshort{hss} when a \acrshort{ue} attaches to the
network. Instead, the PGPP operator issues \acrshort{sim} cards with
\textit{identical} \acrshort{imsi}s to all of its subscribers. In this model,
the \acrshort{imsi} is used only to prove that a user has a valid \acrshort{sim}
card to use the infrastructure and, in turn, the PGPP network can provide an IP
address and connectivity and offer the client a \acrshort{guti}, providing the
user with a unique identity necessary for basic connectivity.

\label{sec:auth}
5G authentication is normally accomplished using \acrshort{imsi}s at the
\acrshort{hss}; however, all PGPP users share a single \acrshort{imsi}. Thus, to
authenticate a user, we designed a post-attach oblivious authentication
scheme to ensure that the PGPP operator is able to account for the user
without knowing who they are.

\paragraph{PGPP Gateway.} In order to perform this authentication we create a new
logical entity called a PGPP Gateway (\acrshort{pgppgw}), shown in Figure~\ref{fig:lte_arch}, which sits between the
\acrshort{pgw} and the public Internet. The \acrshort{pgw} is configured to have
a fixed tunnel to a \acrshort{pgppgw}, which can be located outside of the PGPP
operator's network.  Using this mechanism, the \acrshort{pgppgw} only sees an IP
address, which is typically NATed, and whether that IP address is a valid user.
Notably, it does not have any information about the user's \acrshort{imsi}. The
\acrshort{pgppgw} design also allows for many different architectures. For
instance, multiple \acrshort{pgppgw}s could be placed in multiple datacenters or
even use a privacy service such as Tor.\footnote{We leave exploration into such
scenarios to future work.}

\paragraph{Authentication properties.} From the perspective of the
\acrshort{pgppgw}, there are multiple properties an authentication scheme must
guarantee: (1) the gateway can authenticate that a user is indeed a valid
customer\footnote{Due to ``Know Your Customer'' rules in some jurisdictions, the
provider may need to have a customer list, necessitating that the user
authentication scheme be compatible with periodic explicit customer billing.};
(2) the gateway and/or any other entities cannot determine the user's identity,
and thus cannot link the user's credentials/authentication data with a user
identity; and (3) the gateway can determine whether a user is unique or if two
users are sharing credentials.

%

\begin{table}[t]
\centering
\small
\scalebox{1.0}{
\begin{tabular}{|l||c|c|c|}
\hline
\textbf{Scheme} & \textbf{Customer?} & \textbf{Anonymous?} & \textbf{Unique?}\\\hline\hline
Standard auth & $\bullet$ & & \\\hline
Group/ring sig & $\bullet$ & $\bullet$ & \\\hline
Linkable ring sig & $\bullet$ & & $\bullet$ \\\hline\hline
Cryptocurrency & & $\bullet$ & $\bullet$ \\\hline
PGPP tokens & $\bullet$ & $\bullet$ & $\bullet$ \\\hline
\end{tabular}}
\caption{Three properties needed for user authentication in a privacy-preserving cell network and schemes to achieve them.}
\label{t:gateway-auth}
\vspace{-3.5mm}
\end{table}

As we show in Table~\ref{t:gateway-auth}, the challenge is that standard
approaches for authentication only provide one of the three required properties
and widely-studied cryptographic mechanisms only provide two of the three
properties. For example, an ordinary authentication protocol (of which there are
many~\cite{bellare1993entity, jakobsson2001mutual}) can provide property 1) but
not 2) and 3).  A cryptographic mechanism such as group
signatures~\cite{chaum1991group,boneh2004short} or ring
signatures~\cite{cramer1994proofs,rivest2001leak} can protect the user's
identity upon authentication, providing properties 1) and 2), but not 3) as
providing the last property would violate the security of the signature scheme.
Similarly, traitor tracing schemes~\cite{chor1994tracing} (such as for broadcast
encryption~\cite{fiat1993broadcast}) can provide all three properties but in
practice cannot provide property 3) as the traitor tracing would require actual
physical confiscation of the ``traitor'' phone by the \acrshort{mvno}, which is
infeasible. A variation on ring signatures known as linkable ring
signatures~\cite{liu2004linkable} provides the ability for a user's identity to
be revealed if the user signs multiple messages with the same key. While this is
useful in establishing that the user is unique and hasn't shared their
credentials, it also partially violates the user's anonymity, as that key cannot
be used again.

\paragraph{Effective authentication.} There are two approaches that we view as
viable, depending on the circumstances.  An anonymity-preserving
cryptocurrency can provide properties 2) and 3), but not 1) as a cryptocurrency
would combine billing and authentication at the \acrshort{pgppgw}. For
\acrshort{mvno}s that are not required to know their customers, an
anonymity-preserving cryptocurrency may be the ideal solution for both user
authentication and payment, though even the best coins provide imperfect
anonymity guarantees~\cite{kappos2018empirical}.  

To provide all three properties, we develop a simple scheme called \textbf{PGPP
tokens} that helps us sidestep the issues with alternative approaches.  The
choice of authentication scheme is deployment-context specific. With PGPP
tokens, when paying a monthly bill a user retrieves authentication tokens that
are blind-signed using Chaum's classic
scheme~\cite{chaum1983blind,bellare2003one} by the billing system.  Later, when
authenticating to the service, the user presents tokens and the service (the
\acrshort{pgppgw}) verifies their signature before allowing the user to use the
network.  The token scheme ensures that the service can check the validity of
tokens without identifying the user requesting access.  The user then presents
the next token in advance so as to ensure seamless service. Note that PGPP
tokens disallow the post-pay model for cellular billing, as the network would
be required to know the identity of users in order to accurately charge them
for usage. Therefore, PGPP is pre-pay only, though this can be adjusted to
emulate post-payment (\eg, users pre-pay for tokens on an ongoing
basis rather than only monthly, and tokens are valid for a longer time period, such as a year, rather
than for only one billing period).

Each token represents a unit of access, as is appropriate for the service
provider.  Some providers may choose to offer flat-rate unlimited-data service,
in which case each token represents a fixed period of time; this is the default
approach that we use to describe the scheme below. Other providers may choose
to offer metered service, in which case each token represents a fixed unit of data,
such as 100 MB or 1 GB, rather than a period of time.  Still others may choose
to provide two-tiered service priority by marking each token with a priority bit, in addition to either unlimited data or
metered data service; such prioritization does come with slight privacy loss,
as the \acrshort{mvno} and \acrshort{mno} alike would be able to differentiate which priority level was in use. The
privacy loss of two-tiered data priority can be partially mitigated by
offering all users some amount of time or GB of high-priority service after
which they must fall back to low-priority service; such a service plan
structure is fairly standard in the industry today.  In such a setting, each
user would have both high-priority and low-priority tokens and thus would not
be clearly stratified into two identifiable groups of users.

At the beginning of a billing period, the billing system defines
$s$ time slices (\eg, corresponding to hours) or another unit of access (\eg, a unit of data) and generates $s$ RSA keypairs
for performing blind signatures using Chaum's scheme.  It then appends the
public keys for this time period to a well-known public repository that is
externally maintained (\eg, on GitHub), and these are fetched by users. The user generates $s$ tokens where
each token takes the form $i\|r$ where $i$ is the time slice index as a 256-bit
unsigned value zero indexed from the beginning of the billing period, and $r$ is
a 256-bit random value chosen by the user.  The user then blinds these tokens.
The user pays the bill using a conventional means of payment (\eg, credit
card), and presents the blinded tokens to the billing system to be signed; the
system signs each token with the corresponding time slice key and returns these
values to the user. The user unblinds the response values and verifies the
signatures for each.

Upon later authentication to the service, the user presents its signed token for
the current time slice to the \acrshort{pgppgw}, which verifies the signature
and if valid begins forwarding the user's traffic onto the Internet.  Since the
token signature was generated using Chaum's scheme, the service cannot determine
which human user corresponds to which signed token.  If the same token is used
by two different users during the same time period then the service can conclude
that a user has shared their credentials and is attempting to cheat.

The costs of this scheme to both the PGPP operator and the user are low. The
operator stores the list of used tokens in a standard consistent and replicated
cloud database, so the service can operate multiple \acrshort{pgppgw}s, though
it is likely that a small number of \acrshort{pgppgw}s can serve a large number
of users: we benchmarked the 2048-bit RSA signature verification used here at
31$\mu$s per call using Crypto++~\cite{cryptopp} on a single core of a 2.6GHz
Intel Xeon E5-2640 CPU, and thus with a single CPU core the \acrshort{pgppgw}
can handle token verification for tens of millions of users. The tokens
themselves are small and the storage cost to the provider is about 1.5 MB / user
per time period, which is a small amount for any user's phone to store and for a
provider even hundreds of millions of tokens amounts to mere GBs of data in
cloud storage.

\paragraph{User device agent.} To automate the process of authenticating with the
\acrshort{pgppgw}, we create a simple agent that runs as background job on the
user device. This agent leverages the Android JobScheduler API; in the event of
cellular connectivity, the JobScheduler triggers PGPP-token-based authentication
with the \acrshort{pgppgw}.  The agent establishes a TLS connection to the
\acrshort{pgppgw} and then sends the token for the current time slice. Once the
user presents a valid token, the \acrshort{pgppgw} begins forwarding traffic for
that user, and thus this behavior is akin to a captive portal though the
authentication is automatic and unseen by the user.


\subsection{Location privacy}\label{sec:locationprivacy} As described
in~\cref{sec:location}, cellular operators track user location in the
form of tracking areas for \acrshort{ue}s in order to quickly find users when
there is incoming content. PGPP leverages an existing mechanism in
the cellular standard to reduce the effectiveness of {\em local-targeted}
attacks described in~\cref{sec:attacks}. 

Paging has been exploited in the past to discover user location by adversaries.
However, the use of tracking areas is useful for the cellular provider in that
it confines the signaling message load (\ie, paging messages) to a relatively
small subset of the infrastructure. Tracking areas reduce mobility signaling
from \acrshort{ue}s as they move through the coverage zone of a single tracking
area. Note that emergency calling represents a special case in cellular
networks. When a device dials 911, the phone and network attempt to estimate
accurate location information. In this work we do not alter this functionality
as we anticipate that users dialing 911 are willing to reveal their location. 

\label{sec:tals} In PGPP, we exploit the tracking area list
(\acrshort{tal}) concept, introduced in 3GPP Release 8~\cite{3gpp.23.401}. Using
\acrshort{tal}s, a \acrshort{ue} no longer belongs to a single tracking area,
but rather is given a list of up to 16 tracking areas that it can freely move
through without triggering a tracking area update, essentially creating larger
tracking areas. Whereas prior work has focused on using \acrshort{tal}s to
pre-compute optimal tracking area combinations for
users~\cite{razavi,razavi2,razavi3}, in PGPP, we use \acrshort{tal}s to provide
provide improved location anonymity. Typically, \acrshort{tal}s consist of
groups of adjacent tracking areas that are pre-computed, essentially growing the
tracking area for a \acrshort{ue} to the union of all tracking areas in the
\acrshort{tal}. We do not use \acrshort{tal}s in this way. Instead, we generate
\acrshort{tal}s on-the-fly and generate them uniquely for each \acrshort{ue}.
When a \acrshort{ue} attaches or issues a tracking area update message, the
\acrshort{mme} learns the \acrshort{enb} and tracking area the \acrshort{ue} is
currently attached to. The \acrshort{mme} then generates a unique \acrshort{tal}
by iteratively selecting at random some number (up to the \acrshort{tal} limit
of 16) of additional, adjacent tracking areas. By generating unique
\acrshort{tal}s for each user, attackers are unable to know a priori which set
of tracking areas (or \acrshort{enb}s) that victim is within. We explore
tradeoffs in terms of \acrshort{tal} length, control traffic overhead, and
location anonymity in the next section.

%% file: analysis.tex
\section{Analysis}\label{sec:simulation} To study the implications of a PGPP
deployment, we create a simulation to model users, mobility, and cell
infrastructure. We study the impact of PGPP's design on various cellular attacks
that occur today. We then analyze the inherent tradeoffs from the PGPP
operator's perspective, as improved privacy comes at the price of increased
control traffic. Lastly, we examine PGPP in a lab testbed on real devices.  

\subsection{Simulation configuration}
\textit{\acrshort{enb} dataset.} We select Los Angeles County, California as the
region for our simulation, which provides a mix of both highly urban areas as
well as rural areas. For \acrshort{enb} location information, we use
OpenCellID~\cite{opencellid}, an open database that includes tower locations and
carrier information. To simplify the simulation, we select base stations from
the database that are listed as providing LTE from AT\&T, the provider with the
most LTE eNodeBs (22,437) in the region. We use LTE eNodeBs as the number of
\acrshort{enb}s deployed remains small.

\begin{figure}[t]
\centering
\includegraphics[width=.85\columnwidth]{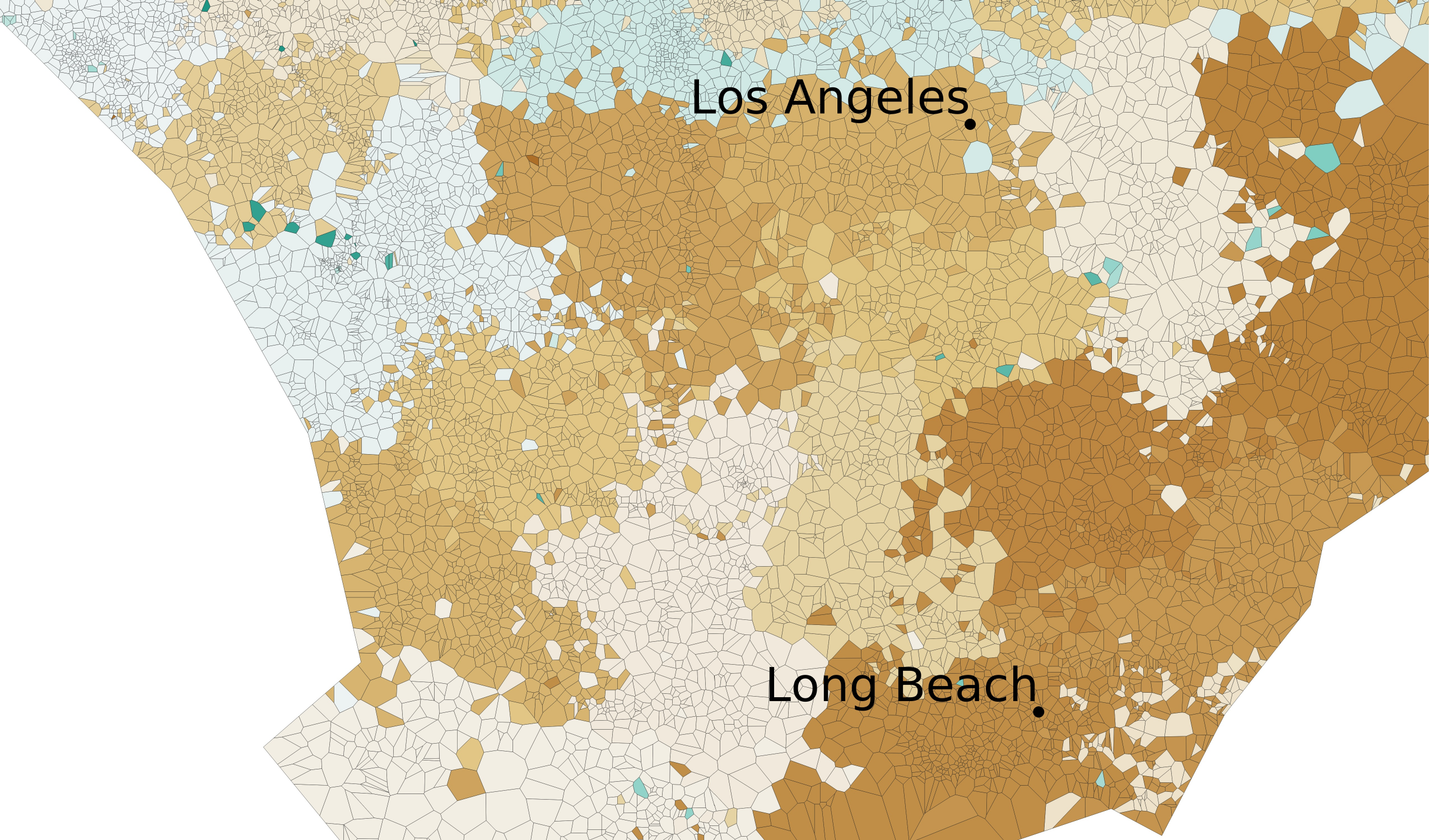}
\caption{Partial simulation map. Cells are shaded by AT\&T tracking area.}
\label{fig:map}
\end{figure}

Given their geographic coordinates, we estimate coverage areas for every
\acrshort{enb} using a Voronoi diagram. During the simulation, a \acrshort{ue}
is assigned to the \acrshort{enb} that corresponds to the region the
\acrshort{ue} is located within. While such discretization is not likely in
reality as \acrshort{ue}s remain associated with an \acrshort{enb} based on
received signal strength, this technique provides us with a tractable mobility
simulation. A partial map of the simulation region is shown in
Figure~\ref{fig:map}. ENodeB regions are shaded based on the tracking area value
in the OpenCellID database.
\begin{figure}[t]
\centering
\includegraphics[width=.75\columnwidth]{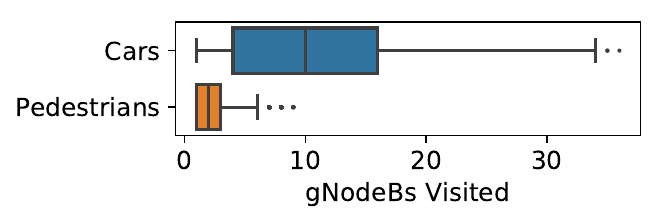}
\caption{gNodeBs visited by simulated mobile users.}
\label{fig:usereNodeBs}
\end{figure}

\textit{Mobility traces.} To simulate realistic mobility patterns (\ie, users
must follow available paths), we generate mobility traces using the Google
Places~\cite{placesapi} and Directions~\cite{directionsapi} APIs. First, we use
the Places API to find locations in the simulation region that are available
when searching for ``post office.'' Each place is associated with latitudinal
and longitudinal coordinates. We then generate mobility traces by randomly
selecting start and end points, and use the Directions API to
obtain a polyline with coordinates along with estimated times to reach points
along the line. We generate 50,000 mobility traces: 25,000 cars and 25,000
pedestrians. We then use ns-3 to process the mobility traces and generate
coordinates for each trace at 5-second intervals, in a method similar
to~\cite{routesmobility}. We use this output, along with the \acrshort{enb}
Voronoi diagram to assign each simulated \acrshort{ue} to an \acrshort{enb} for
every 5-second interval in the mobility trace. Figure~\ref{fig:usereNodeBs}
shows the distribution of the number of \acrshort{enb}s visited by
\acrshort{ue}s in the simulation. As expected, car trips result in a
significantly higher number of \acrshort{enb}s for a \acrshort{ue} compared with
pedestrian trips.

\textit{Synthetic traffic.}\label{sec:traffic} We simulate one hour. To create
control traffic, at every 5-second interval we randomly select 5\% of the user
population to receive a ``call.'' A call results in a paging message that is
sent to all \acrshort{enb}s in the \acrshort{ue}'s tracking area. Each paged
user enters a 3-minute ``call'' if it is not already in one, at which point
further paging messages are suppressed for that user until the call is complete.
We run the simulation with PGPP enabled as well as with the conventional
infrastructure setup.

\textit{Custom \acrshort{ta}s.}\label{sec:customtas} As we detail further
in~\cref{sec:control}, large \acrshort{tal}s increase control traffic loads,
which lowers the network's user capacity. Therefore, we generate new tracking
areas in the underlying network in order to mitigate the control traffic burden.
As tracking areas normally consist of groups of adjacent \acrshort{enb}s, we
need a method by which we can cluster nearby \acrshort{enb}s into logical
groupings. To do so, we use k-means clustering with the \acrshort{enb}
geographic coordinates allowing for Euclidean distance to be calculated between
\acrshort{enb}s. We generate several underlying tracking area maps, with the
number of \acrshort{ta}s (\ie, k-means centers) ranging from 25 to 1,000. For
comparison, the AT\&T LTE network in the simulation is composed of 113
\acrshort{ta}s. 

\subsection{Cellular privacy attack analysis}\label{sec:attackanalysis} Given
the taxonomy we presented in~\cref{sec:attacks}, we analyze the identity and
location privacy benefits of PGPP in the simulated environment.

\paragraph{Global-bulk attacks.}\label{sec:globalbulk} By nullifying the
value of \acrshort{imsi}s, separating authentication with connectivity,
and increasing the broadcast domain for users, we increase user identity privacy
even with an adversary that is capable of bulk surveillance over an entire
network (\eg, operators, governments). 

\textit{Anonymity analysis}
We measure the anonymity of a user when under bulk attacks using \textit{degree
of anonymity}~\cite{10.5555/1765299.1765304}. The degree of anonymity value
ranges from zero to one, with ideal anonymity being one, meaning the user could
be any member of the population with equal probability. In this case, we
consider the \acrshort{imsi} value to be the target identity. The size of the anonymity set
for a population of $ N $ users will result in a maximum entropy of:

\begin{equation}
H_{M}= log_{2}(N)
\end{equation}

The degree of anonymity is determined based on the size of the subset of user
identities $ S $ that an attacker could possibly believe the victim to be: 
\begin{equation}
d = \frac{H(X)}{H_{M}} = \frac{log_{2}(S)}{log_{2}(N)}
\end{equation}

\begin{figure}[t]
	\begin{subfigure}[b]{0.485\columnwidth}
		\centering
		\includegraphics[width=\columnwidth]{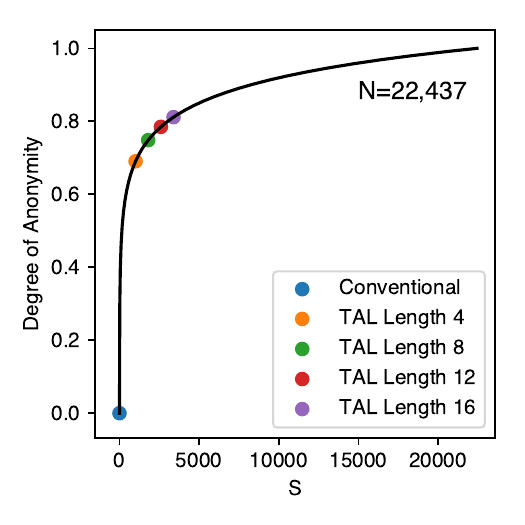}
		\caption{TALs.}
		\label{fig:talanon}
    \end{subfigure}
    \hfill
    \begin{subfigure}[b]{0.485\columnwidth}
		\centering
		\includegraphics[width=\columnwidth]{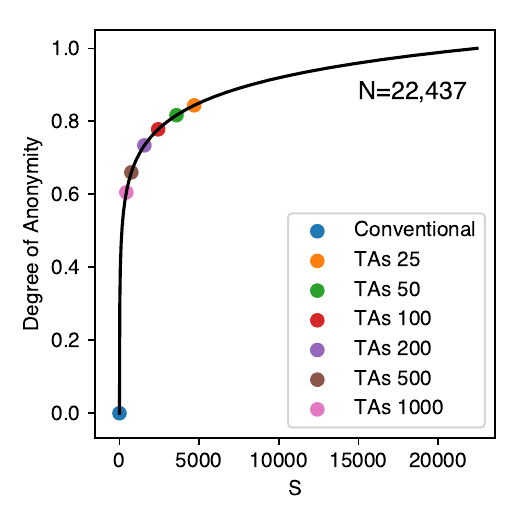}
		\caption{Custom TAs.}
		\label{fig:customanon}
    \end{subfigure}
    \caption{Degree of anonymity using TALs and custom TAs.}
    \vspace{-3mm}
\end{figure} 

Given global visibility into the network, we can reason about the anonymity set
using the number of \acrshort{enb}s that a victim could possibly be connected
to. This is because a cellular carrier can know the exact base station that a
user is connected to once the \acrshort{ue} enters an active state. As a
baseline, the anonymity set for traditional cellular is $
\frac{log_{2}(1)}{log_{2}(22,437)} = 0 $, as each \acrshort{imsi} is a unique
value. With PGPP, \acrshort{imsi}s are identical, so from the perspective of the
carrier, the victim could be connected to any \acrshort{enb} that has at least
one PGPP client connected to it. Using our simulated environment we collect, for
each paging message, the number of \acrshort{enb}s that had users within their
range and use the median value to calculate the degree of anonymity.
Figures~\ref{fig:talanon} and~\ref{fig:customanon} show the degree of anonymity
using different configurations of \acrshort{tal}s and custom \acrshort{ta}s,
respectively. We see that high degrees of anonymity are attainable despite an
attacker's global visibility. For instance, with \acrshort{tal}s of length 8,
the degree of anonymity is 0.748. 

\paragraph{Local-bulk attacks.}\label{sec:localbulk} PGPP's use of
identical \acrshort{imsi}s reduces the importance of \acrshort{imsi}s, and by
extension the usefulness of local bulk attacks on user identity. An attacker
that can view traffic at the \acrshort{enb}(s) can gain insight into nearby
\acrshort{imsi}s. 

In traditional cell networks, each user has a globally unique \acrshort{imsi}
($S = 1$), resulting in a degree of anonymity of zero as the victim could only
be one user. In our measurement study (\cref{sec:zaatarimeasurement}), we showed
that \acrshort{imsi}s are routinely broadcast over cell networks, making an
\acrshort{imsi} catcher or SDR attack powerful. The subset $ S $ in PGPP, on the
other hand, is the size of the population of PGPP users in a given location, as
all \acrshort{imsi} values are identical and a local bulk attacker cannot know
the true identity of a single user. To get an idea of $ S $, we can calculate
the number of PGPP users connected to each \acrshort{enb} in the simulation.
Over the course of the simulation, we find a mean value of 223.09 users
connected to each \acrshort{enb} that has users, which results in a degree of
anonymity $ \frac{log_{2}(223.09)}{log_{2}(50,000)} = 0.50 $. While this value
is somewhat low compared to the ideal value of $ 1 $, it is a drastic improvement over
conventional cellular architecture, and is dependent on the overall user
population in the network. As more PGPP users exist, the degree of anonymity
increases.

\paragraph{Local-targeted attacks.}\label{sec:localtargeted} In PGPP,
local-targeted attacks to discover a user's location are diminished in two ways:
first, \acrshort{imsi}s are no longer a useful ID, so identifying an individual
among all users is challenging; and second, we use \acrshort{tal}s to increase
the paging broadcast domain for a given \acrshort{ue}. From an attacker's point
of view, this broadens the scope of where the target \acrshort{ue} may be
located. 

\begin{figure}[t]
	\begin{subfigure}[b]{0.485\columnwidth}
		\centering
		\includegraphics[width=\columnwidth]{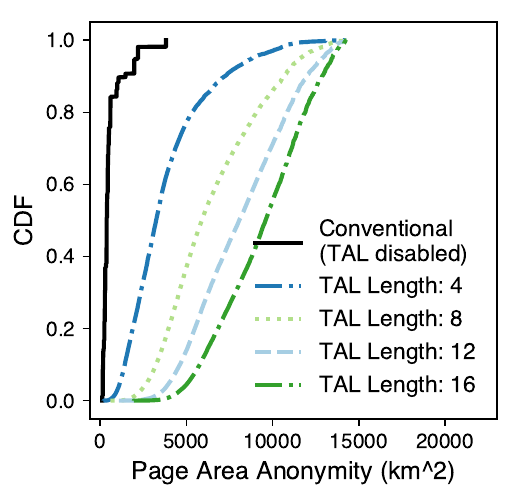}
		\caption{TALs.}
		\label{fig:areas}
    \end{subfigure}
    \hfill
    \begin{subfigure}[b]{0.485\columnwidth}
		\centering
		\includegraphics[width=\columnwidth]{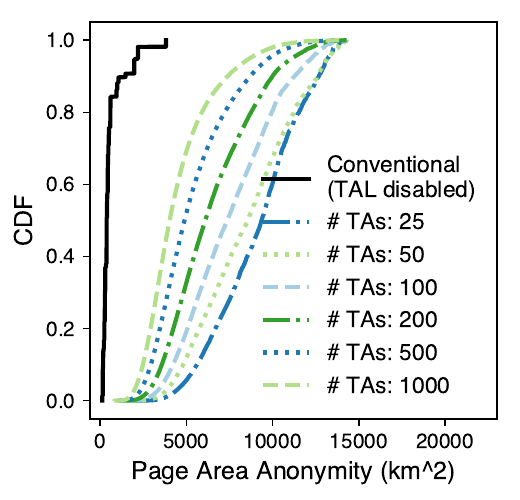}
		\caption{Custom TAs.}
		\label{fig:areacustom}
    \end{subfigure}
    \caption{Area anonymity using TALs and custom TAs.}
    \vspace{-3mm}
\end{figure} 

In Figure~\ref{fig:areas}, we plot the CDF of geographic areas in which pages
are broadcast as we increase \acrshort{tal} lengths using the base map
consisting of 113 tracking areas. We calculate the area by generating a bounding
box around all \acrshort{enb}s that are included in the broadcast domain. As
shown, large \acrshort{tal}s result in drastically higher area anonymity
compared with \acrshort{tal}s disabled, particularly considering the number of
\acrshort{ue}s that could potentially be located in the larger geographic areas.
For instance, the median area for the conventional simulation is 378.09
km\textsuperscript{2} whereas \acrshort{tal} lengths of 8 and 16 result in
median areas of 5,876.96 and 9,585.17 km\textsuperscript{2}, respectively.

We analyze anonymity with \acrshort{tal}s of length 16 while
the underlying map is varied using custom \acrshort{ta}s. Figure~\ref{fig:areacustom} shows our results. We
observe that as the number of tracking areas increase, resulting in smaller
tracking areas, the area anonymity decreases. However, despite the decrease, the
area anonymity remains considerably larger than anonymity with \acrshort{tal}s
disabled as \acrshort{tal}s include additional tracking
areas. For instance, the median area for the conventional case is 378.09
km\textsuperscript{2} whereas the median area for a base map of 500 tracking
areas with \acrshort{tal} 16 is 4891.08 km\textsuperscript{2}, a nearly 13-fold
increase from the perspective of a local targeted attacker.

\subsection{Impact of PGPP on network capacity}\label{sec:control} From an
operational perspective, the privacy benefits delivered by PGPP must coincide
with feasibility in terms of control overhead in order for it to be deployable.
Control traffic determines network capacity in terms of the number of users that
are serviceable in a given area. In this
section, we explore control traffic load when using \acrshort{tal}s.

\subsubsection{Control overhead with PGPP TALs}
We first seek to quantify control message overhead while we leverage tracking
area lists to provide location anonymity against local-targeted attacks. Recall from~\cref{sec:tals} that we randomly
select additional tracking areas from the simulated coverage area to create
\acrshort{tal}s, which increases the broadcast domain for a page. Increased
control traffic impacts both \acrshort{enb}s and \acrshort{mme}s, however, from
our experience with real cellular networks the control traffic capacity at
\acrshort{enb}s is the bottleneck as \acrshort{mme}s have much higher capacity.
Thus, we focus on \acrshort{enb} control load.  

\begin{figure}
	\centering
	\begin{subfigure}[b]{0.485\columnwidth}
		\centering
		\includegraphics[width=\columnwidth]{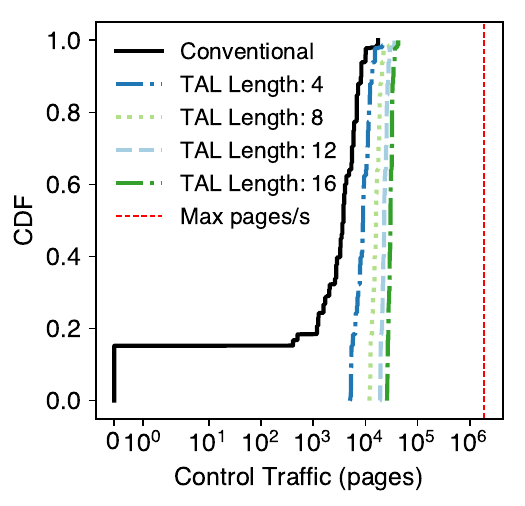}
		\caption{Control traffic with TALs.}
		\label{fig:control}
	\end{subfigure}
	\hfill
	\begin{subfigure}[b]{0.485\columnwidth}
		\centering
		\includegraphics[width=\columnwidth]{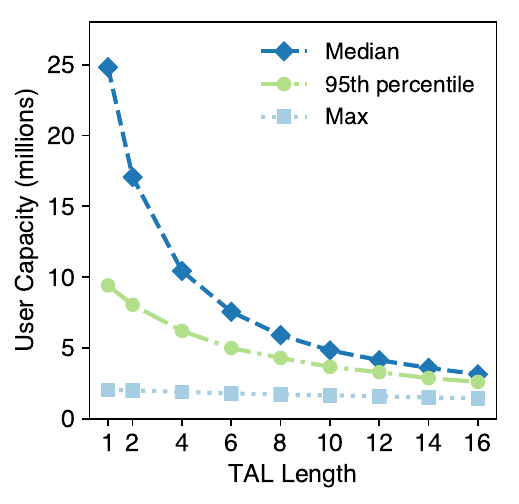}
		\caption{Capacity with TALs.}
		\label{fig:users}
	\end{subfigure}
	\caption{Control traffic and system capacities leveraging PGPP TALs in the simulated environment.}
\end{figure}

Figure~\ref{fig:control} shows a cumulative distribution function (CDF) for the
number of pages broadcast by the simulated \acrshort{enb}s. In the figure,
``Conventional'' corresponds to disabling \acrshort{tal} functionality. As
expected, larger \acrshort{tal} lengths result in increased control traffic for
\acrshort{enb}s as they are more likely to be included in the paging broadcast
domain for a given \acrshort{ue}. 

To gain insight into the control limitations of real \acrshort{enb}s, we
consider the capabilities of a Huawei BTS3202E eNodeB~\cite{huawei-enodeb}, which is limited to 750 pages per second.
When capacity planning, it is commonplace to budget paging traffic headroom;
accordingly, we estimate the maximum paging capacity for an \acrshort{enb} to be
525 pages per second (70\% of the BTS3202E capacity). This value is depicted in
the vertical red line in the figure (525 pages $\times$ 3600 seconds = 1,890,000
pages/hour). The simulation allows us to illustrate the user population that
could be supported by the network, provided a population with similar mobility
and traffic profiles as defined in~\cref{sec:traffic}. Recall that we simulate
50,000 users, both pedestrians and cars. We consider the paging load for the
network and select the \acrshort{enb}s with the maximum paging load, the 95th
percentile, and the median to estimate the number of users each could
theoretically support by taking into account the max page limitation of the
BS3202E. Figure~\ref{fig:users} shows the user capacity as \acrshort{tal}
lengths are increased. A \acrshort{tal} length of one shows the conventional
network, as the \acrshort{tal} is composed of a single tracking area. As expected,
larger \acrshort{tal}s result in a reduction in the number of users the
\acrshort{enb}s can handle compared with performance when \acrshort{tal}s are
disabled, due to increased paging load.

\begin{figure}
	\centering
	\begin{subfigure}[b]{0.485\columnwidth}
		\centering
		\includegraphics[width=\columnwidth]{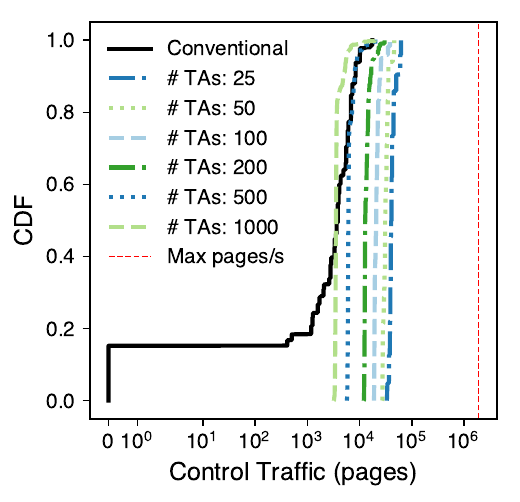}
		\caption{Custom TAs: Control traffic.}
		\label{fig:controlcustom}
	\end{subfigure}
	\hfill
	\begin{subfigure}[b]{0.485\columnwidth}
		\centering
		\includegraphics[width=\columnwidth]{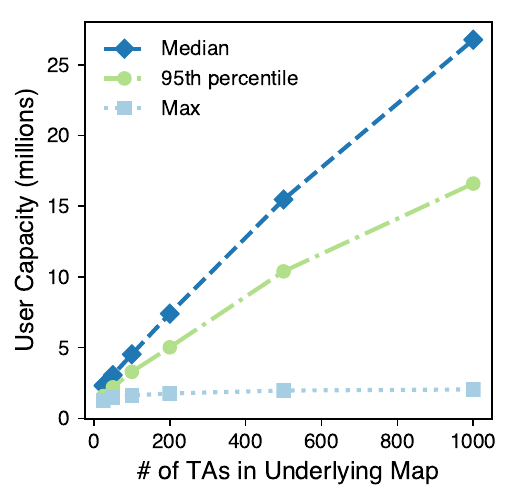}
		\caption{Custom TAs: Capacity.}
		\label{fig:userscustom}
	\end{subfigure}
	\caption{Control traffic and system capacities with custom tracking areas in the simulated environment.}
\end{figure}

\subsubsection{Control overhead with custom tracking areas}
As  we've demonstrated, large \acrshort{tal}s result in
\acrshort{enb}s with higher control traffic load, effectively reducing the user
capacity the network. To explore whether we can re-gain control traffic we again
consider new, custom tracking area maps that are generated using k-means where
we vary the number of unique tracking areas in the simulated network.

We run the simulation with various custom tracking area maps, with all
\acrshort{ue}s using \acrshort{tal} lengths of 16. The results are shown in
Figures~\ref{fig:controlcustom} and~\ref{fig:userscustom}. We observe that a
basemap consisting of 25 tracking areas leads to even higher control traffic
compared with the conventional (\ie, AT\&T) tracking area map. A map consisting
of more tracking areas results in \acrshort{ta}s with fewer \acrshort{enb}s,
thus reducing the paging load. We see that a map of 500 \acrshort{ta}s, even
with a \acrshort{tal} of length 16, results in similar paging load compared with
the conventional map with \acrshort{tal} disabled. Correspondingly, the user
capacity of the network with a higher number of tracking areas nears the
conventional capacity from Figure~\ref{fig:users}.

\subsection{Testbed analysis}
We study our PGPP design on a lab testbed in order to understand potential
drawbacks. We implement a software-based \acrshort{EPC} and connect commodity
phones to the software-defined radio-based \acrshort{enb}.

\begin{figure}[t]
\centering
\includegraphics[width=0.9\columnwidth]{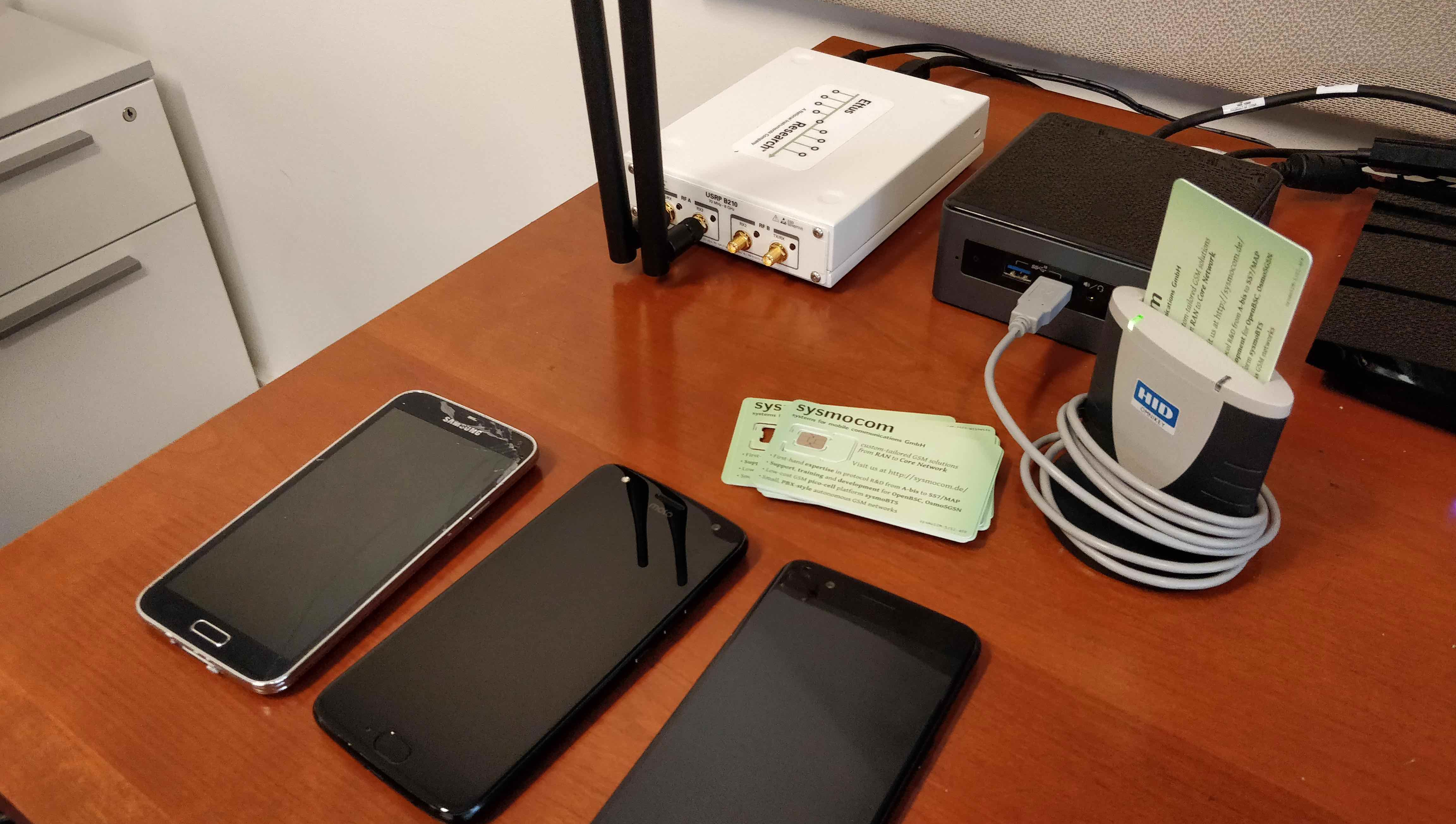}
\caption{PGPP prototype test hardware.}
\label{f:pgppgear}
\vspace{-2mm}
\end{figure}

\paragraph{Prototype.}\label{sec:prototype} We create our prototype code on
srsLTE~\cite{srsLTE}, an open-source platform that implements LTE-compliant base
station and core network functionality and can be run using software-defined
radios\footnote{We build our prototype on a 4G LTE platform as we are not aware
of any platforms that fully implement 5G and are sufficiently
mature for experimentation with real hardware.}. Our testbed, shown in
Figure~\ref{f:pgppgear}, consists of an Intel Core i7 machine running Linux and
a USRP B210 radio. We use off-the-shelf commodity phones (Moto X4, Samsung
Galaxy S6, and two OnePlus 5s) with programmable \acrshort{sim} cards installed
to allow the phones to connect to the PGPP network. 

SrsLTE maintains contexts for each connected \acrshort{ue} related to mobility
and connectivity. The contexts are stored as structs that include the
\acrshort{ue} \acrshort{imsi} in a simple key-value store, with the
\acrshort{imsi} serving as the key. When the \acrshort{mme} receives
mobility-related messages, it checks against the appropriate contexts to handle
the requests. We add an additional value, a PGPPIMSI, into the context structs.
The PGPPIMSI is generated by combining the \acrshort{imsi} with a temporary
value that is unique to the individual
\acrshort{ue}-\acrshort{enb}-\acrshort{mme} connection. Accordingly, each
\acrshort{ue} has a unique PGPPIMSI, which then allows us to look up the correct
context when managing states.

\paragraph{Identical IMSIs and Shared Keys.}
Given identical \acrshort{imsi} values for all users, the PGPP attach procedure
can result in additional steps compared with the traditional attach. This is
caused by sequence number synchronization checks during the authentication and
key agreement (\acrshort{aka}) procedure, which is designed to allow the
\acrshort{ue} and the network to authenticate each other. The fundamental issue
is that the \acrshort{hss} and the \acrshort{sim} maintain a sequence number
(\acrshort{sqn}) value that both entities increment with each successful attach.
As multiple devices use the same \acrshort{imsi}s, the sequence numbers held at
the \acrshort{hss} and on individual devices will no longer match, causing an
authentication failure (known as a sync\_failure). At that point the
\acrshort{ue} re-synchronizes with the \acrshort{hss}. 

\begin{figure}[t]
\centering
\includegraphics[width=0.9\columnwidth, trim={0 10 0 13},clip]{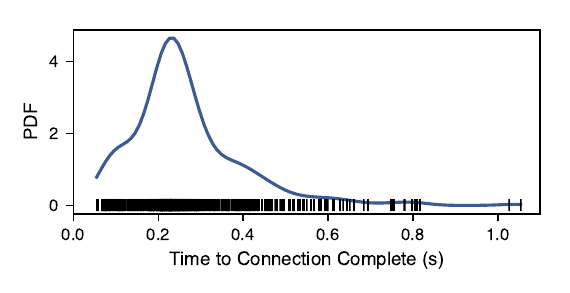}
\caption{Connection delays due to sync\_failure.}
\label{f:sqnconnect}
\end{figure}

We explore the delay introduced by sync\_failures using our testbed.
Figure~\ref{f:sqnconnect} shows a PDF of the delays to connection completion for
\acrshort{ue}s that hold identical \acrshort{imsi}s and attempt to authenticate
simultaneously. In order to trigger many simultaneous authentication requests,
we use openairinterface5G~\cite{nikaein2014openairinterface} to create 100
simulated \acrshort{ue}s. We observe in that the first successful \acrshort{ue}
usually takes roughly 200 ms to connect, while subsequent \acrshort{ue}s that
experienced sync\_failures experience additional delays. In our relatively small
experiment the \acrshort{ue}s all successfully connect to the network within 1.1
seconds. In a large-scale production network the number of UEs that
simultaneously attempt to connect would be larger. PGPP-based networks can
mitigate the issue by using more \acrshort{hss}es, which would reduce the number
of \acrshort{ue}s that each \acrshort{hss} is responsible for. Fortunately, the
push for 5G will lend itself to many \acrshort{hss}es as the core network
entities are being redesigned to be virtualized and located nearer to
\acrshort{ue}s.

%% file: discussion.tex
\section{Related Work}
Prior work on anonymous communications often traded off latency and
anonymity~\cite{vandenHooff:2015:VSP:2815400.2815417,199303,corrigan2010dissent,Corrigan-Gibbs:2015:RAM:2867539.2867658}.
Likewise, Tor~\cite{dingledine2004tor} and Mixnets~\cite{chaum1981untraceable}
also result in increased latency while improving anonymity. However, such
solutions are inappropriate for cellular systems as, apart from SMS, cellular
use cases require low latency. Additionally, the architecture continues
to utilize identifiers (\eg, \acrshort{imsi}) that can expose the user to
\acrshort{imsi} catcher attack or allow for location tracking by the operator. 

There has been extensive prior work on finding security and
privacy issues in cellular
networks~\cite{4215735,8835335,practicalattackslte,lteinspector,kune2012location}.
We decouple the \acrshort{imsi} from the subscriber by setting it to a single
value for all users of the network. Altering the \acrshort{imsi} to specifically
thwart \acrshort{imsi} catcher and similar passive attacks has been previously
proposed~\cite{vandenBroek:2015:DIC:2810103.2813615,
Khan:2017:TIC:3098243.3098248,sung2014location,Arapinisnew}. These techniques
use pseudo-\acrshort{imsi}s (PMSIs), which are kept synchronized between the
\acrshort{sim} and the \acrshort{hss}, or hypothetical virtual \acrshort{sim}s,
allowing for user identification. We aim to go beyond thwarting \acrshort{imsi}
catchers, and do so while considering active attacks without requiring
fundamental changes on the \acrshort{ue}; we protect users from the operator
itself. 

Hussain \etal{} introduce the TORPEDO attack~\cite{hussain2019privacy}, which
allows attackers to identify the page frame index and using that, the presence
or absence of a victim in a paging broadcast area (\ie, a tracking area).
However, our use of tracking area lists to provide additional paging anonymity
(\cref{sec:locationprivacy}) increases the location in which a victim could
potentially be, reducing the effectiveness of third-party paging-related
localization attacks. The authors also define the PIERCER attack, which enables
the attacker to reveal a victim's \acrshort{imsi} with only their phone number.
PGPP nullifies this attack by making all \acrshort{imsi}s identical. Cellular
signaling protocols have been demonstrated by multiple works to leave users'
privacy vulnerable to
attack~\cite{lorenz2001securing,sengar2006ss7,engel2008locating,holtmans2016detach,sonar}.
Our initial design avoids signaling protocol vulnerabilities by providing data-only rather than
voice/SMS, and roaming to other networks can be enabled by requiring
home-routing rather than local breakout. Hussain \etal{} identifies a 5G vulnerability that
allows an attacker to neutralize \acrshort{guti} refreshment
in~\cite{Hussain:2019:PSP:3319535.3354263}. However, this requires a MiTM attack
(\eg, \acrshort{imsi} catcher), which necessarily means the attacker knows the
victim's location. Additionally, the \acrshort{guti} is a temporary identifier,
and is not associated with a specific user.

Choudhury and K\o ien alter \acrshort{imsi} values, however both require
substantial changes to network
entities~\cite{Choudhury:2012:EUI:2360018.2360115,6673421}. We argue that a
privacy-preserving architecture must be fully compatible with existing
infrastructure as the global telecom infrastructure is truly a
network of networks, comprised of multiple operators that connect via well-known
APIs.  

\section{Concluding Remarks}
User privacy is a hotly contested topic today, especially as law enforcement
organizations, particularly in authoritarian states, insist upon increasingly
ubiquitous surveillance.  In addition, law enforcement has long demanded
backdoor access to private user devices and user data~\cite{savage2018lawful}. 

We do not believe that users of PGPP, in its current form, would be capable of
withstanding targeted legal or extra-legal attacks by nation-state organizations
(\eg, the FBI or NSA), though PGPP would likely limit the
ability of such organizations to continue to operate a regime of mass
surveillance of user mobility.  In addition, a more common and problematic form
of privacy loss today is due to the surreptitious sale of user data by network
providers; this is a matter PGPP addresses in a manner that aligns
with user autonomy.  Our aim is to improve privacy in line with prior societal
norms and user expectations, and to present an approach in which
privacy-enhanced service can be seamlessly deployed.